\documentclass[%
reprint,
 amsmath,amssymb,
 aps,
]{revtex4-1}

\usepackage{graphicx}
\usepackage{dcolumn}
\usepackage{bm}

\usepackage{subfigure}  
\usepackage{placeins}
\usepackage{multirow}
\usepackage{rotating}

\begin{document}


\title{Epidemic prevalence information on social networks mediates emergent 
collective outcomes in voluntary vaccine schemes}

\author{Anupama Sharma$^{1}$, Shakti N. Menon$^{1}$, V. Sasidevan$^{1,2}$ \& Sitabhra Sinha$^{1}$}
\affiliation{${}^{1}$The~Institute~of~Mathematical~Sciences,~CIT~Campus,~Taramani,~Chennai~600113,~India. \\
${}^{2}$Department of Physics, Cochin University of Science and
Technology, Cochin 682022, India.}

%
%
%

\date{\today}

\begin{abstract}
The success of a 
vaccination program is crucially dependent on its adoption by a
critical fraction of the population, as the resulting herd immunity
prevents future outbreaks of an epidemic. However, the effectiveness
of a campaign can engender its own undoing if individuals choose to
not get vaccinated in the belief that they are protected by herd
immunity. Although this may appear to be an optimal decision, based on a rational
appraisal of cost and benefits to the individual, it exposes the population to
subsequent outbreaks. We investigate if voluntary vaccination can
emerge in a an integrated model of an epidemic spreading on a social
network of rational agents that make informed decisions whether to be
vaccinated. The information available to each agent includes the
prevalence of the disease in their local network neighborhood and/or
globally in the population, as well as the fraction of their neighbors
that are protected against the disease. Crucially, the payoffs
governing the decision of agents evolve with disease prevalence,
resulting in the co-evolution of vaccine uptake behavior with the
spread of the contagion. The collective behavior of the agents responding
to local prevalence can lead to a significant reduction in the final
epidemic size, particularly for less contagious diseases having low
basic reproduction number $R_0$. Near the epidemic threshold
($R_0\approx1$) the use of local prevalence information can result in
a dichotomous response in final vaccine coverage. The implications of
our results suggest the nature of information used by individuals is a
critical factor determining the success of public health intervention
schemes that involve mass vaccination.\\
\end{abstract}

\maketitle


\section{\label{sec:level1} Introduction}

\par Immunization through the vaccination of populations has been
estimated to annually prevent 2-3 million deaths from infectious
diseases such as measles, diphtheria, pertussis and
tetanus~\cite{who2016}. This number may rise
substantially with the development of strategies to further increase global vaccine
coverage~\cite{CDC2016}. Apart from conferring a long-term protection 
against the disease to the vaccinated individual, 
vaccination has an even more important community-level benefit. 
A sufficiently high vaccine coverage makes it difficult for the
pathogen to find susceptible hosts, thereby conferring \emph{herd
immunity} to the whole population~\cite{And1985,Hee2015}. Consequently, even those
members of the community who are unable to get vaccinated, such as
newborns and immune-suppressed individuals, are protected against the
disease. In principle, any disease caused by a pathogen that only has
human hosts can be eradicated by mass immunization. Such an outcome
has been realized for smallpox~\cite{Fen1988, Ore2017} and is expected to be achieved
for polio~\cite{Lar2011a,gvap}. 
Conversely, the presence of a significant fraction
of non-immunized individuals, which disrupts
the population's herd immunity, can result in the recurrent outbreaks of
vaccine-preventable diseases such as measles, mumps and pertussis~\cite{CDC2013}.
Elucidating the mechanisms 
that promote wider acceptance of vaccination 
in the population can therefore help explicate the reasons behind the failure of
immunization programs. 

\par One of the most important challenges in implementing an effective
immunization program is to ensure that enough individuals
voluntarily decide to vaccinate. This decision could be based on many
factors such as an individual's knowledge about the costs, including perceived
side-effects, and benefits of vaccination, as well as the social, economic and
cultural environment to which they belong~\cite{Figu2016,Wes2017}.
The lack of public confidence in the efficacy and/or safety of vaccines 
can give rise to vaccine hesitancy (i.e., delay or refusal to get 
vaccinated despite the availability of vaccine services)~\cite{Lar2014}, and in extreme
cases generate vaccine scares~\cite{Bla2011,Lar2016}.  Even in the absence of 
any bias against a vaccine as such, vaccine uptake in the population may 
vary over time with changing prevalence of the disease. Indeed, it is expected 
that individuals will be more likely to get themselves vaccinated when there is a higher
risk of getting infected~\cite{Hor2015}. Conversely, low disease incidence may often
lead to a significant drop in vaccine uptake, presumably because of
the lower perceived risk of contracting the disease~\cite{Jan2003}.
This suggests that when the threat of infection is high the individual
has a strong incentive to get vaccinated, while at times of lower risk
she may be tempted to avoid vaccination and free-ride on the herd immunity 
provided by immunized members of a population without bearing any cost 
herself. However, if everyone argues in this manner and avoids vaccination, 
it would leave the population completely exposed to invasion by the pathogen. 
This is essentially an instance of a {\em social dilemma}~\cite{Feh2003} that often 
arises in strategic interactions between rational individuals. That is, while free-riding
appears to be optimal from an individual's perspective, it leads to a 
clearly undesirable collective outcome. This is one of the 
problems central to game theory, which therefore provides a natural 
framework for understanding the conditions under which a
population of rational individuals will voluntarily decide to get
vaccinated.

\par 
Earlier studies of interaction between disease spreading and vaccine
uptake behavior 
that incorporated a game theoretic framework have typically assumed
homogeneous, well-mixed populations~\cite{Bau2003,Bau2004,Bau2005}. Thus, 
the risk of infection for every individual, as well as the
protection offered to them by immunized individuals in their
neighborhood, is identical. However, in reality, individuals interact
primarily with neighboring members of their social networks
and can have widely different contact structures~\cite{Wasser1994}.
Considering the network microstructure governing contacts between
individuals can explain aspects of the collective
outcomes of spreading contagion processes~\cite{Barrat2008} and strategic
interactions~\cite{Szabo2007} that do not manifest in well-mixed models of
populations.
However, models that investigate vaccine uptake behavior by individuals
in social networks typically
do not incorporate strategic considerations in terms of explicit
payoffs, i.e., the net benefit associated with specific collective
actions.
Instead, agents are assumed to imitate the
behavior of their more ``successful'' neighbor~\cite{Fen2011}. 
As models incorporating strategic decision-making and those utilizing
social network approaches each describe different aspects of
vaccine uptake behavior~\cite{Wan2016}, a framework
combining both may come closer at capturing the complexity
associated with such behavior in reality.

\par 
To understand the interaction between human behavior and epidemic
dynamics~\cite{Fer2007,Fun2010}, in this paper we present a model in which rational
agents take strategic decisions to vaccinate themselves on the basis of
information about the disease prevalence and the immune status of
their neighbors on a social network. Each agent decides their action
by playing a game against 
a hypothetical opponent who shares the same neighborhood as it.
Unlike previous studies that use a similar framework of strategic
interactions, in our model the payoffs
defining the structure of the game 
incorporate
real-time information on the specific situation
prevailing in the network neighborhood
and consequently vary dynamically amongst individuals.
Thus, the games played by the different agents co-evolve with the
spread of the disease across the network, resulting in an emergent
spatio-temporal heterogeneity in the nature of the games.
We find that this heterogeneity at the level of individual agents can have
significant implications for population-level outcomes such as the final
epidemic size and the extent of vaccine coverage. 
We also examine how the source of the information, viz., global (fraction of the population that is infected) or local (fraction of infected
neighbors), that agents may use in
assessing the risk of getting infected can lead to 
very different collective outcomes.
The implications of our results reported here suggest that
access to real-time information about the state of an
evolving epidemic can change the risk perception and affect the
vaccine uptake decisions taken by individuals. These in turn result in
emergent patterns of collective choice behavior that may provide
useful insights into the mechanisms driving vaccine acceptance, 
which could be relevant for public health planning.

\section{\label{sec:level2} Model and Methods} 
\par  In our model, we study the dynamics of two coupled processes,
namely epidemic spreading and the evolution of vaccine uptake
behavior, on a social network of $N$ agents. The connection topology
of the network is specified by the contact structure among individuals
in a given population. 
The time-scale of an epidemic considered here is much
shorter than durations over which the network structure 
may change significantly as a result of births, deaths and
migrations of individuals.
The spread of the disease over the network
changes the status of an agent which, at any instant, can
be in one of three possible states,
namely, susceptible ($S$), infected ($I$), and recovered ($R$). We assume
that recovery from the disease confers immunity from further infection 
to an agent. 
The disease is assumed to
spread through direct contact between agents with a
transmission probability $\beta$, while infected agents recover from the disease after an
average time period of $\tau_{I}$ .
Thus, the disease dynamics follows the well-known SIR
model~\cite{May1991}. Introducing {\em vaccination} in this framework
allows a susceptible individual to avoid the possibility of getting
infected by immediately achieving an immune status (which effectively
corresponds to the $R$ state).

As an epidemic propagates through the population,
each agent can have access to {\em local information} about the 
number of infected cases among her network neighbors (i.e., with whom
she has direct contact), as well as {\em global information}
about the disease prevalence in the entire network.
In reality, such information is obtained through different channels,
e.g., via mass-media in the case of global information and through
word of mouth for local information.
The agents also have information about the extent to which their neighborhood
offers them protection from the disease. 
This is provided by their
knowledge of how many of their neighbors are immune as a result of
either having recovered from the disease earlier, or through
vaccination.
Each agent utilizes the above information to determine their
likelihood of getting infected. Based on this threat perception, the
agents subsequently make a strategic decision on whether to get
vaccinated by taking into account the ``cost'' associated with vaccination.
This cost arises from the threat of side-effects, either real or
perceived, as well as the effort involved in getting vaccinated,
and tempts the agent to free-ride on the protection that may be
offered by the immunity of their neighbors.
By engaging in such behavior agents can enjoy the benefits of
immunization without bearing
the cost of getting vaccinated themselves.
However, if every agent argues along the same lines, it will lead to
extremely low vaccine uptake, causing the loss of herd
immunity and exposing the population to the risk of an epidemic
outbreak of a vaccine-preventable disease.
This results in a dilemma for a population of well-informed rational
agents.

\begin{figure}
\includegraphics[width=8.5cm]{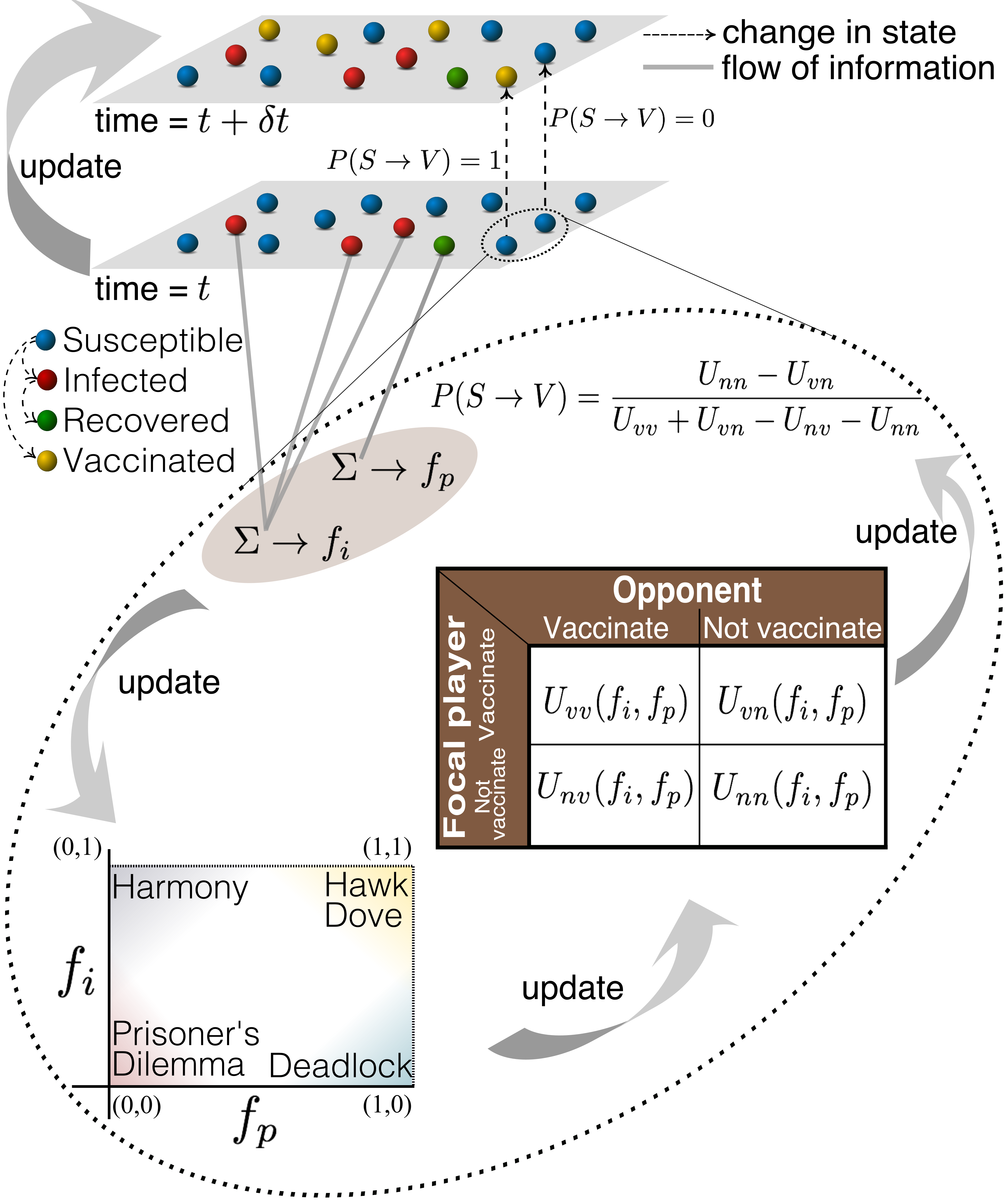}\\
\caption{\label{fig:00}  Schematic representation of the coupling between the spread of an epidemic and strategic vaccine uptake behavior by individuals. Agents are classified according to their state with respect the disease as Susceptible (S), Infected (I), Recovered (R) and Vaccinated (V). The two layers represent the states of the nodes at two time instants. The broken lines represent the change in the state of agents and grey solid lines represent the flow of information about the state (infected and removed) of agents in the network. The curved arrow between the two layers represents the update (time evolution) of the system. The broken curve encloses the game-theoretic process that determines whether an agent decides to vaccinate or not, based on the probability of an agent choosing to get vaccinated $P(S \rightarrow V)$. The table inside the broken curve is a payoff matrix used by an agent to make decisions. Here the ``opponent'' is a hypothetical agent having identical information, choices of actions and associated payoffs. The payoff received by the focal player is represented by a function of the form $U_{xy}(f_{i},f_{p})$, where $x$ and $y$ are, the actions of the focal player and opponent respectively. The fractions of infected and protected (immune) agents are represented by $f_{i}$ and $f_{p}$, respectively. By varying these two parameters the nature of the game can change between different classes, as shown in the inset to the lower left.}
\end{figure}

As a game-theoretic framework provides a natural setting for
investigating such social dilemmas, we model the vaccine uptake decision
process of individual agents in terms of games.
In order to make a strategic decision each agent plays a symmetric
2-person game against
a hypothetical opponent who shares the same neighborhood and hence has
identical information.
At each round of the game, an agent has a choice of two possible
actions, i.e., to get vaccinated ({\em v}) or not ({\em n}).
The cost and benefit associated with the choices is represented 
in terms of a payoff matrix.
An important feature of our approach is that
the payoffs evolve with the progress of the epidemic and the ensuing
change in vaccine coverage in the population.
The payoff received by the focal player $j$, where $j \in [1,N]$, is represented by a function
of the form $U_{xy}(f_{i},f_{p})$, where $x,y \in \{n,v\}$ are the actions
of the focal player and the virtual opponent, respectively (see table
in Fig.~\ref{fig:00}).
Here, $f_p$ is the fraction of neighbors that are immune
and $f_i$ is a linear
combination of local and global information about the disease
prevalence:
\[f_i(j)=\underbrace{\alpha (I/N)}_{\rm
global}+\underbrace{(1-\alpha)(k_{i}(j)/k(j))}_{\rm local}.\]
Note that $I$ is the number of infected agents in a population of size
$N$, while $k(j)$ is the total number of neighbors of the focal agent $j$, of which
$k_i(j)$ individuals are infected. 
By tuning the parameter $\alpha \in [0,1]$, we can consider any
information scenario between the two extreme cases wherein an agent
uses exclusively
local ($\alpha=0$) or global ($\alpha=1$) information.

\par 
 As mentioned earlier, a high disease prevalence (i.e., large value of $f_i$) will 
act as an incentive for the focal agent to get vaccinated. Therefore, 
as the benefits of vaccination outweigh its cost when prevalence is high, 
it is reasonable to assume that the values of the payoffs
associated with the decision to \emph{vaccinate} increase
with the increase in prevalence.  
In other words, $U_{vv}(f_i,f_p)$ and
$U_{vn}(f_i,f_p)$ are increasing functions of $f_{i}$. Additionally,
when all her neighbors are immune to the disease,
there is a high probability that the focal agent will successfully
escape infection even without vaccination. 
Hence, we assume that values of the payoffs
associated with the decision to \emph{not vaccinate} increase with the
increase in the fraction of protected neighbors. This suggests that
$U_{nv}(f_i,f_p)$ and $U_{nn}(f_i,f_p)$ are increasing functions of
$f_{p}$.
 For concreteness, we choose the simplest possible linear functional form for   
$U_{nv}$,
$U_{nn}$, $U_{vv}$ and $U_{vn}$ as follows: 
\begin{align*}
&U_{nv}= a f_p + b, 
&U_{nn}= c f_p + d,\\
&U_{vn}= e f_i + f,
&U_{vv}= g f_i + h.
\end{align*}
This linear form in $f_i$ and $f_p$ has the added advantage of not having 
multiple solutions (i.e., Nash equilibria, explained later) for any particular choice of $f_i$ and $f_p$, which would have required 
invoking additional selection criteria for choosing among them.    
As the payoff functions are time-varying, the nature of the game can change depending on the 
hierarchical relation between the payoffs that prevails at any instant.  
\par To characterize the hierarchy of payoff functions in the $(f_p,f_i)$ space, we note that when $f_i$ is high and $f_p \rightarrow 1$, it is possible to escape infection as long as most of the neighbors are immune but in the absence of protection from the neighborhood, vaccination is vital to an individual. This suggests the following relation between payoffs: $U_{nv}>U_{vv}>U_{vn}>U_{nn}$, i.e., the game is Hawk-Dove~\cite{HD}. When $f_i$ is low and $f_p \rightarrow 1$, the non-vaccinators prevail as there a very low risk of infection and most of the population is immune to the disease. This would result in $U_{nv}>U_{nn}>U_{vv}>U_{vn}$, i.e., the game is Deadlock~\cite{Dead}. When $f_i$ is high and $f_p \rightarrow 0$, the benefits of vaccination outweigh the perceived cost of vaccination because of the high risk of contracting disease. This results in the hierarchal relation $U_{vv}>U_{vn}>U_{nv}>U_{nn}$, i.e., the game is Harmony~\cite{Har}. When $f_i$ is low and $f_p \rightarrow 0$, it is extremely tempting to not get vaccinated because of low prevalence. However the possibility of being infected is non-zero, which makes vaccination a viable choice. This results in $U_{nv}>U_{vv}>U_{nn}>U_{vn}$, i.e., the game is Prisoner's Dilemma~\cite{PD} (see Fig.~\ref{fig:00}, inset). 
These four games govern the preference that an agent has for each action (viz., to vaccinate or to not vaccinate) at the four extremities of the $(f_{i},f_{p})$ parameter space. In the interior of this space, the hierarchies among the payoffs gradually change, thereby giving rise to different games. To ascertain that the system behaves in the same way as explained above at these four extremities, we choose the parameters $a-h$ such that $U_{nv}$, $U_{nn}$, $U_{vn}$ and $U_{vv}$ satisfies the inequalities mentioned above. The payoff associated with not getting vaccinated when the opponent chooses to vaccinate $(U_{nv})$ is always greater than the corresponding payoff for the case where of both do not get vaccinated $(U_{nn})$, as the latter situation exposes both to the risk of being infected. We hence set $a=c$ without loss of generality. Similarly, the payoff received when both players get vaccinated $(U_{vv})$ is always greater than that obtained when only the focal player is vaccinated vaccinated $(U_{vn})$ as she alone bears the cost associated with vaccination. We hence set $e=g$ without loss of generality. If the parameters a-h satisfy the following relation:\[ a+b>e+h>e+f>b,a+d>h>d>f, \]   
then the situations discussed above (Hawk-Dove, Deadlock, Harmony and Prisoners' Dilemma) will prevail at the four extremities of the $(f_{i},f_{p})$ space.

\par As the epidemic spreads in the population each susceptible agent $j$ will, at any time $t$, choose an action such that a unilateral change of action will not yield a higher payoff. In game theory, such an action profile is known as a Nash equilibrium~\cite{Hol2004}. If player $j$ (and her opponent) decides to vaccinate with probability $p_j$ ($p_o$) and not vaccinate with probability $1-p_j$ ($1-p_o$), 
the expected utility for agent $j$ can then be calculated as
\begin{align*}
\epsilon_{j} = p_j(p_o(U_{vv}+U_{nn}-U_{nv}-&U_{vn})+U_{vn}-U_{nn})\\
&+p_o(U_{vn}-U_{nn})+U_{nn}.
\end{align*}
Given that the game is symmetric, the Nash equilibrium 
would be either $p_j=0$ or $p_j=1$ if it is pure, or if it is mixed then the agent $j$ would vaccinate with the probability 
 \[p_j=\frac{U_{nn}-U_{vn}}{U_{vv}+U_{nn}-U_{nv}-U_{vn}}.\] 
Note that the expression for the vaccination probability for a mixed strategy Nash equilibrium is similar to the strategy referred as mixed ESS in the Bishop-Cannings theorem~\cite{MSim1982}.
As this probability will be different for each susceptible agent, it introduces heterogeneity in the decision making process over the network.  Also, as this probability can change with time, an agent can change her decision as the disease spreads over the network. Incorporating such spatio-temporally varying strategies for the vaccine uptake of agents on a network presents a more realistic way of examining the coupled dynamics of vaccination and disease.    

 \par In order to study the consequences of the interplay between the strategic decision-making process for vaccine uptake and epidemic spreading, we simulate the stochastic spread of a directly transmitted disease on empirical social networks of villages in southern India~\cite{Ban2013}, as well as model networks. All agents in our model are initially susceptible and  $0.5\%$ of the nodes in a network are randomly chosen to become infected to simulate the onset of an epidemic. Note that no node is initially in a vaccinated state. We employ the Gillespie stochastic evolution algorithm~\cite{Gill1977} to determine the time at which the next event will happen and which node would take part in that event. The event could be one of the three different types of transitions that can change the state of a node: (i) disease transmission~($S \rightarrow I$), (ii) recovery~($I \rightarrow R$), and (iii) vaccination~($S \rightarrow R$). Disease transmission is a contact-dependent transition and can take place only when node $j$ in state $S$ is in contact (i.e., has a connecting link) with nodes in state $I$. Recovery is a time-dependent transition and depends on the time interval spent by a node $j$ in infected state (for more details see~\cite{Jas2016}). Vaccination is an information-driven transition, which involves strategic decision making (as shown in Fig.~\ref{fig:00}). The simulation is stopped when there are no infected nodes remaining in the network.


\begin{figure}
\begin{subfigure}
        \centering
	\includegraphics[width=4cm]{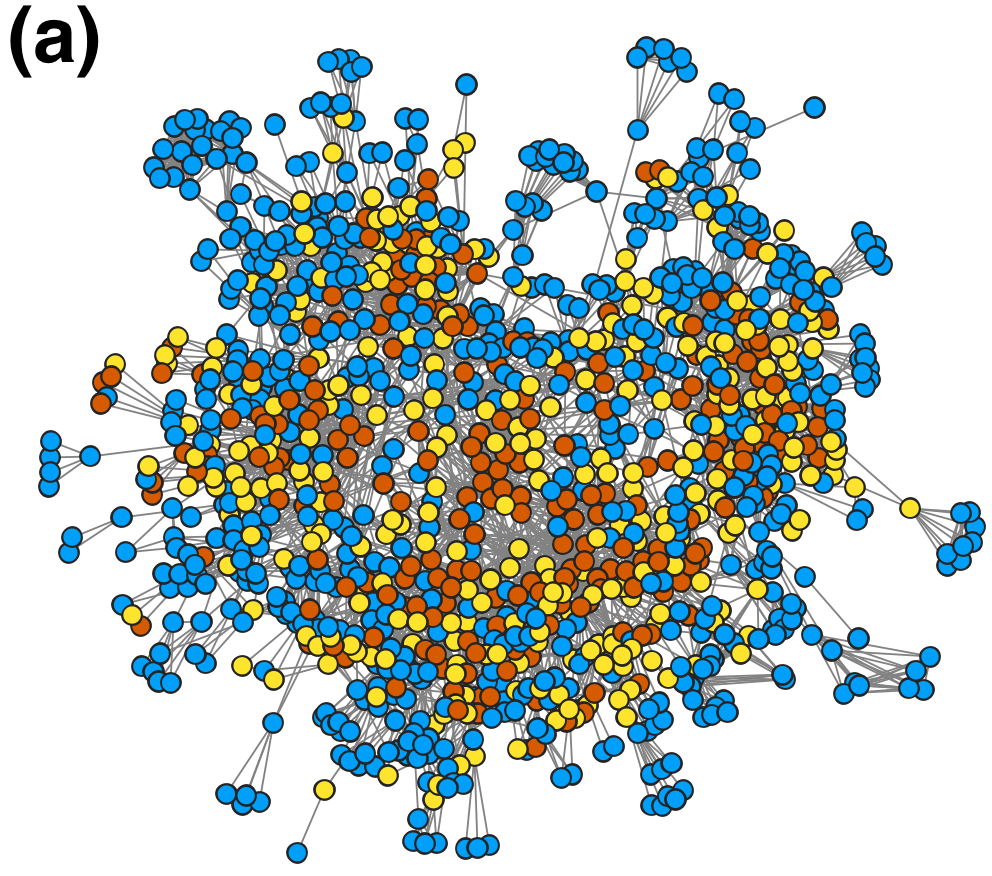}
\end{subfigure}
\begin{subfigure}
 	\centering
	\includegraphics[width=4cm]{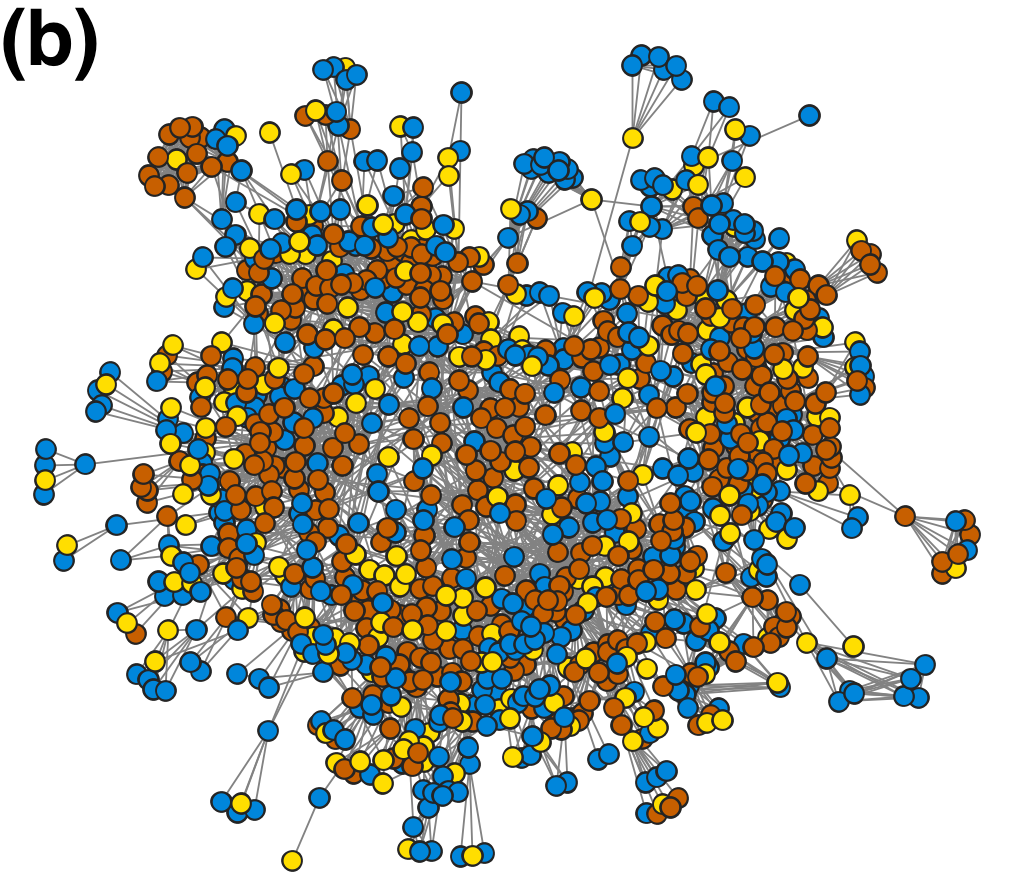}
\end{subfigure}
\includegraphics[width=8.5cm]{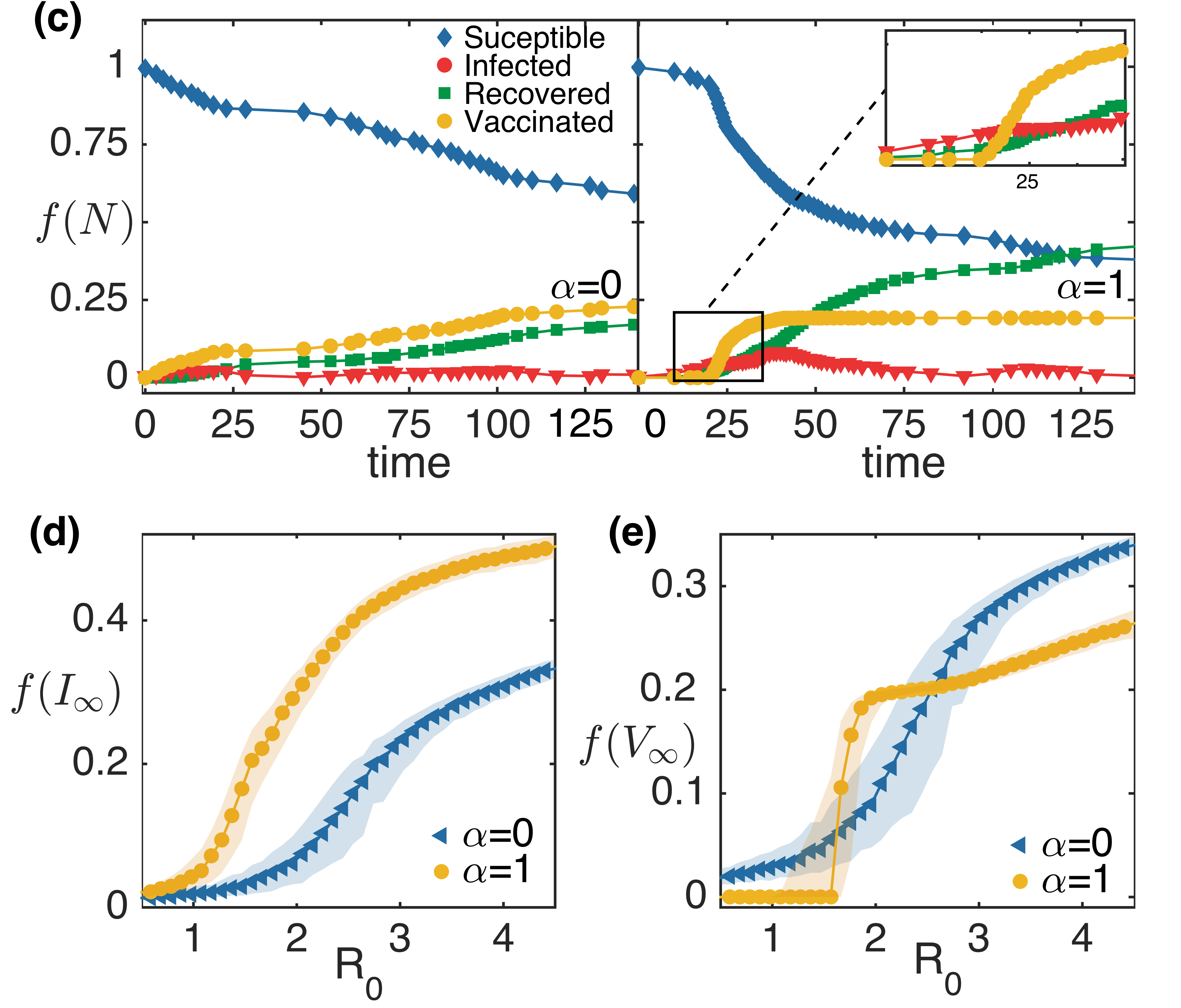}
\caption{\label{fig:01} Simulation results for the co-evolution of epidemic spreading and vaccine uptake behavior in the largest connected component of a social network in a village of southern India~\cite{Ban2013} (village 55~\cite{Jas2016}) with $N=1180$ and $\langle k \rangle=9.78$. A snapshot of the network showing the final states of nodes following a simulated epidemic with  $\beta=0.025$ and $\tau_I=10$ is shown for (a) $\alpha=0$  and (b) $\alpha=1$. The colors of the nodes are representative of the final state: blue, susceptible; yellow, vaccinated; red, recovered (i.e., infected during the epidemic). (c) A sample time series showing the evolution of $S, I, R$ and $V$ for a simulated epidemic with $\beta=0.025$ and $\tau_I=10$ for $\alpha=0$ (left) and $\alpha=1$ (right).  The inset of (c) provides a closer view of the sudden emergence of vaccination when the prevalence becomes sufficiently high. A comparison of the final fraction of agents (d) infected $f(I_{\infty})$ and (e) vaccinated $f(V_{\infty})$ during a simulated epidemic with different values of $R_0$, for $\alpha=0$ and $\alpha=1$. Each of the points represents the median of 1000 simulation runs and the patches indicate the interquartile range (IQR).}
\end{figure}

\begin{figure}
\includegraphics[width=8.5cm]{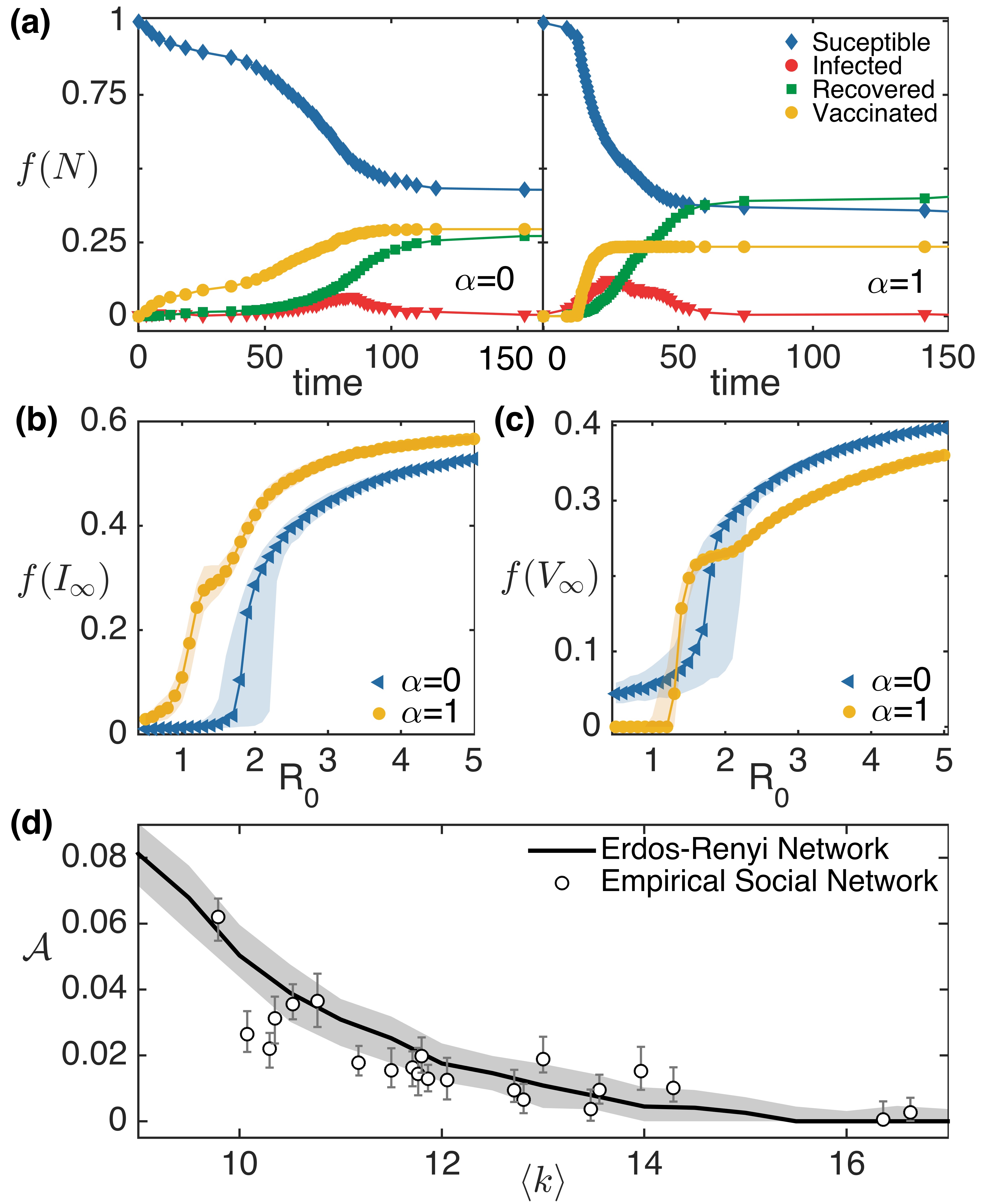}
\caption{\label{fig:02} Simulation results for the co-evolution of epidemic spreading and vaccine uptake behavior in Erd\H{o}s-R\'enyi networks with $N=1024$ and $\langle k \rangle=10$. (a) A sample time series showing the evolution of $S, I, R$ and $V$ for a simulated epidemic with $\beta=0.02$ and $\tau_I=10$ is displayed for $\alpha=0$ (left) and $\alpha=1$ (right).  We display a comparison of the final fraction of agents that are (b) infected $f(I_{\infty})$ and (c) vaccinated $f(V_{\infty})$ during a simulated epidemic with different values of $R_0$, for $\alpha=0$ and $\alpha=1$. Each point represents the median of 1000 simulation runs and the patches indicate the corresponding IQR. (c)  Dependence of crossover area $\mathcal{A}$ on average node degree $\langle k \rangle$ behaves similarly in a empirical social networks and model random networks. The solid line and patch shows the median and IQR of the 1000 simulated epidemics on Erd\H{o}s-R\'enyi networks respectively. The circle and error bars represent the median and IQR of the 1000 simulated epidemic on social network of villages in southern India that have a largest connected component greater than 1000.}
\end{figure}

\begin{figure}
\includegraphics[width=8.5cm]{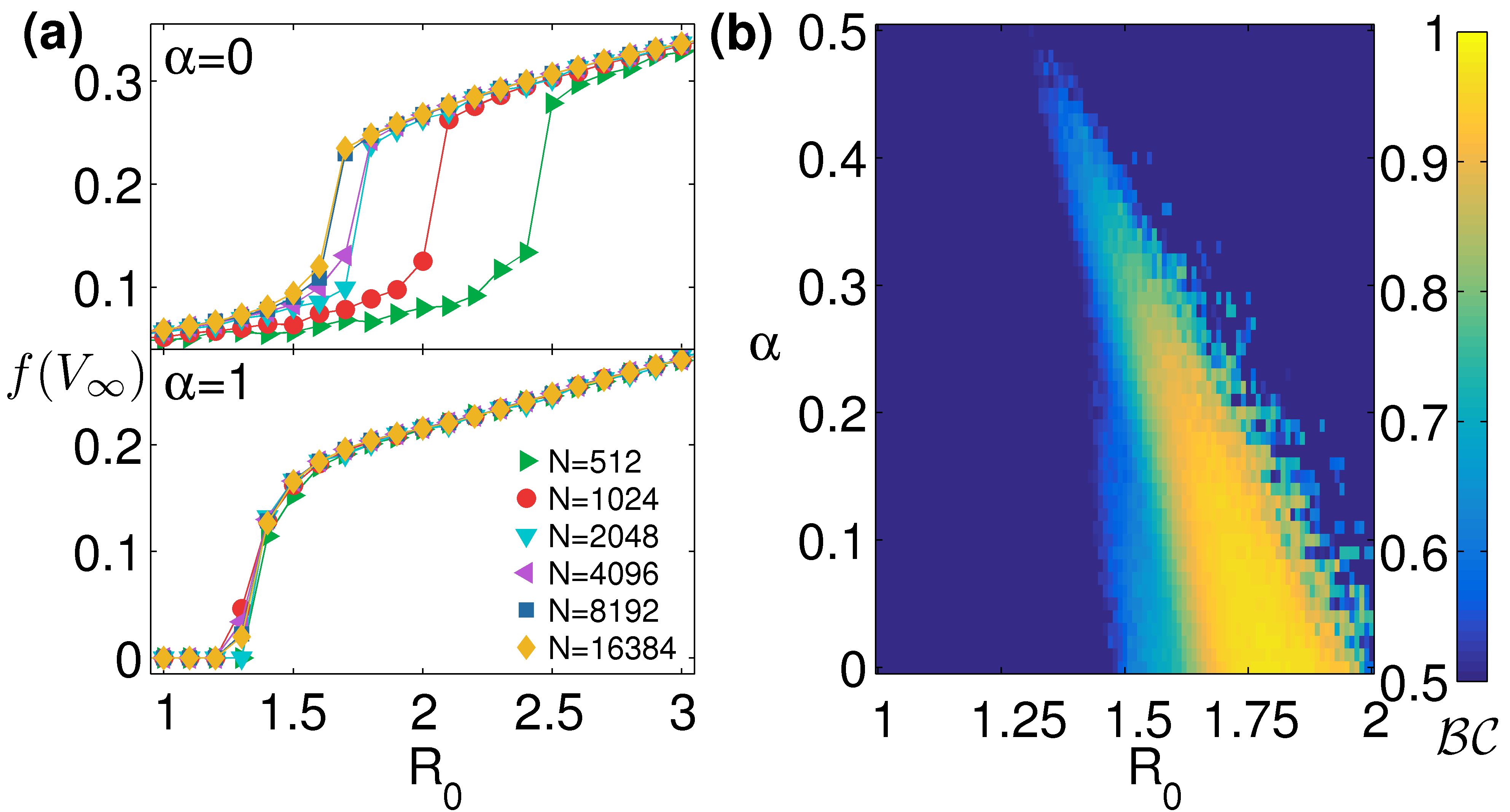}\\
\includegraphics[width=8.5cm]{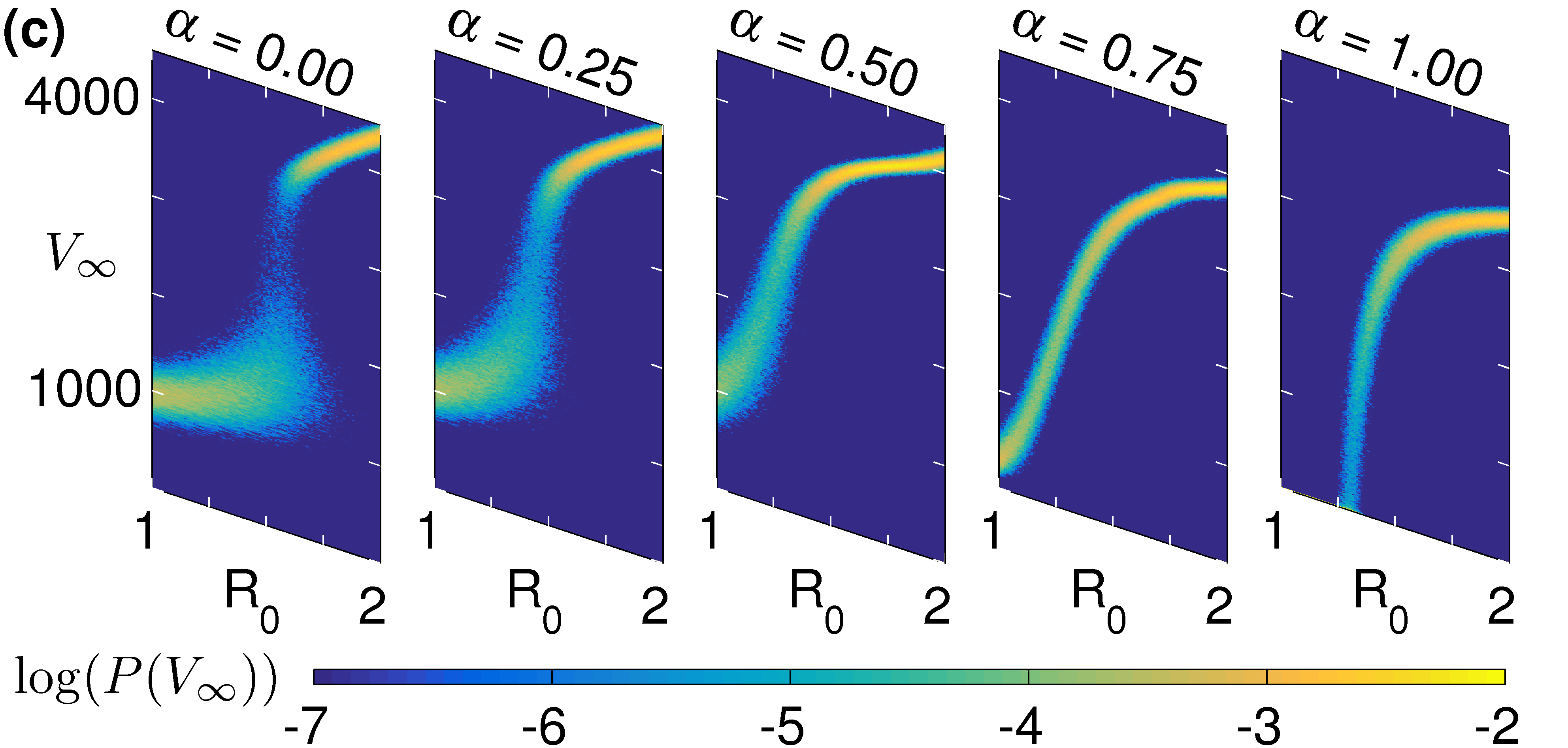}
\caption{\label{fig:03} (a) Assessing the dependence of $f(V_{\infty})$ in a network having average node degree $\langle k \rangle =10$ on population size $N$. The results are shown for  $\alpha=0$ (top) and $\alpha=1$ (bottom). (b)  Bimodality coefficient $\mathcal{BC}$ for the probability distribution of $V_{\infty}$ calculated over $2000$ trials for Erd\H{o}s-R\'enyi networks with $N=16384$ and $\langle k \rangle=10$, and shown over the range of values of $R_0$ and $\alpha$. (c) Probability distribution of $V_{\infty}$  as a function of $R_0$, calculated over $2000$ trials, and shown for different values of $\alpha$.}
\end{figure}


\section{\label{sec:level3} Results}
\par The goal of our model is to see whether voluntary vaccination can emerge as a result of strategic decision making in response to an epidemic threat and if so, what role the source of information (local and global) may play in shaping this response. Fig.~\ref{fig:01} shows the results obtained for a simulated epidemic on the social network of one the 75 villages in southern India from the data set of~\cite{Ban2013}. We stress, however that our results are 
qualitatively similar for other choices of social network (as shown in the subsequent figure). Fig.~\ref{fig:01}(a-b) illustrates the final outcome of a simulated epidemic with transmission probability $\beta=0.025$ and average infectious period $\tau_I=10$ on the empirical social network of a specific village (village $55$ in the data set), for two different values of $\alpha$. The blue color represents the nodes that escaped infection without getting vaccinated.  Note that as all nodes were initially susceptible, the vaccine uptake behavior is entirely epidemic-driven. It is evident from the figure that more agents experience the disease (as indicated by red colored nodes)  when the information available about prevalence is global ($\alpha=1$) as compared to when it is local ($\alpha=0$), although the vaccine coverage (as indicated by yellow colored nodes) is almost same. To understand the reason behind this disparity in the final outcome of epidemic simulated when considering different sources of information,  we consider the time evolution of the fraction of nodes in different states, as shown in Fig.~\ref{fig:01}(c). For $\alpha=0$, the final fraction of agents that were infected during the epidemic, $f(I_{\infty})$, is $0.17$ and the final fraction of agents vaccinated during the epidemic, $f(V_{\infty})$,  is $0.22$. In contrast, for the case $\alpha=1$, $f(I_{\infty})$ is $0.42$ and $f(V_{\infty})$  is $0.19$. Hence, even though $f(V_{\infty})$ is similar for the two cases, there is a significant difference in the value of $f(I_{\infty})$.  It is clear from the figure that voluntary vaccination behavior emerges much later in the case $\alpha=1$ (at $t=20$) as compared to $\alpha=0$, where it emerges almost immediately after initiating the simulated epidemic. As highlighted in the inset of the right panel of  Fig.~\ref{fig:01}(c), in the case $\alpha=1$ the agents start getting vaccinated when the epidemic prevalence becomes significantly high. This emergent behavior is a reasonable description of how responses to epidemics typically unfold. For instance, the media usually reports an outbreak only when the reported cases of the disease becomes sufficiently high. Once a disease has affected a significant proportion of population, even a subsequent high vaccine coverage would be unable to reduce the final fraction of infected agents. To test the robustness of these results with regard to the contagiousness of the epidemic, we simulated the epidemic with different values of $R_0$. For each value of $R_0$ we run $1000$ trials to average over the effect of noise on the final size of the epidemic and vaccine coverage.  On comparing the final outcome of these simulations, it is apparent that the value of $f(I_{\infty})$ for $\alpha=1$ is always greater than the corresponding value for $\alpha=0$, independent of any choice of $R_0$ (Fig.~\ref{fig:01}(d)). This underpins the previous observation that the epidemic infects a larger proportion of agents in the network when agents decide to get vaccinated based on the information about the global disease prevalence, as compared to local. However, a comparison of $f(V_{\infty})$ for $\alpha=0$ and $\alpha=1$ reveals a more complex situation (see Fig.~\ref{fig:01} e). For both low and high values of $R_0$, $f(V_{\infty})$ is higher for $\alpha=0$ than for $\alpha=1$, but there is an intermediate range of $R_0$ in which the values of $f(V_{\infty})$ for $\alpha=1$ are higher than for $\alpha=0$. Thus, there is a crossover of both the curves of $f(V_{\infty})$ for $\alpha=0$ and $\alpha=1$. This shows that an epidemic simulated with these intermediate values of $R_0$ results in higher vaccine coverage when agents base their vaccination coverage on the global information as compared to local. An important point to note here is that the effect of high vaccine coverage in this regime of $R_0$ for $\alpha=1$ is not reflected in the final size of the epidemic (Fig.~\ref{fig:01}(d)). This shows that even if the vaccine coverage in this regime of $R_0$ is high, the simulated epidemic affects more agents for $\alpha=1$ than for the case $\alpha=0$. A possible explanation of this is that in the case of global information the threat perception does not appear significant unless a large proportion of agents are affected by the epidemic and hence fails to overcome the perceived cost of vaccination. This results in limited vaccine uptake which does not provide any significant check on the spread of the epidemic.   
          
\par To gain more insight into the dynamics of the model, we simulated epidemics on Erd\H{o}s-R\'enyi networks with $N=1024$ and average degree $\langle k \rangle=10$, for both $\alpha=0$ and $\alpha=1$ (see Fig.~\ref{fig:02}(a-c)).  The results are consistent with those obtained for the empirical social network. To see how the crossover behavior near epidemic threshold depends on the average degree $\langle k \rangle$ of the network, we simulated the epidemic on Erd\H{o}s-R\'enyi networks having different average degree and on empirical social networks from the dataset of~\cite{Ban2013} with $LCC>1000$. We calculated the area $\mathcal{A}$ enclosed between the two $f(V_{\infty})$ vs $R_0$ curves for $\alpha=0$ and $\alpha=1$. In Fig.~\ref{fig:02}(d), we have shown how this area decreases with an increase in the value of $\langle k \rangle$. This indicates that this intriguing behavior is dependent on the average degree of network.      

\par In order to examine how the results are affected by the size of the population being considered, we display the dependence of $f(V_{\infty})$ on $N$ for $\alpha=0$ (top) and $\alpha=1$ (bottom) in Fig.~\ref{fig:03}(a). We observe that the change in the values of $f(I_{\infty})$ and $f(V_{\infty})$ with respect to $R_0$ show similar behavior on increasing the size $N$ of the network. The  $f(V_{\infty})$ versus $R_0$ curves for these two different values of $\alpha$ show two different kinds of behavior, on increasing the system size. To investigate this change in behavior, we looked into the probability distribution of $f(V_{\infty})$ calculated over $2000$ trials. We found that for $\alpha=1$ this distribution is unimodal for all values of $R_0$, whereas for $\alpha=0$ a bimodal distribution is observed for some values of $R_0$, i.e. the probability distribution has peaks at two different locations. To identify where this behavior changes in the $(R_0, \alpha)$ parameter space, we computed the value of the Bimodality coefficient  $\mathcal{BC}$~\cite{Pfi2013} for the probability distribution of $V_{\infty}$ [shown in Fig.~\ref{fig:03}(b)]. It suggests that the probability distribution of $V_{\infty}$ is bimodal for values of $\alpha <0.5$. This can be observed from Fig.~\ref{fig:03}(c), which shows how the probability distribution of $f(V_{\infty})$ changes on increasing the value of $\alpha$ from $0$ to $1$. This can be a potential signature of a subcritical (discontinuous) transition for local information and a supercritical (continuous) transition for global information.

\section{\label{sec:level4} Discussion}
\par Vaccine hesitancy typically rises with decreasing disease incidence as a consequence of reduced risk perception among individuals of contracting the disease. Understanding the mechanisms driving such behavior is important as it can reverse the success of  any immunization program close to achieving the eradication of a disease~\cite{Saint2013}. We utilize the framework of game theory to investigate vaccine uptake behavior, as it provides an intuitive description for the action of rational agents, i.e. in absence of any social or religious bias against decision to get vaccinated. In contrast to previous approaches, we simulate the spread of an infectious disease on a social network, where each agent can, at every time step, decide whether to get vaccinated. The decision-process of each agent is modelled by a game, in which the payoffs for different actions vary over time as the epidemic progress and the immunization status of the neighboring agents change. Each agent plays against a hypothetical opponent who shares the same neighborhood and thus has identical information, imposing symmetry on the payoff matrix. We examined whether information about an epidemic outbreak at the local or global level can lead to the emergence of voluntary vaccine uptake behavior in a population of agents that are aware of the benefits of free-riding on the immunity of their peers.  In particular, we focused on how spatio-temporal heterogeneity in individuals' vaccine uptake decisions can affect the overall vaccine coverage at the population level, and consequently determine the fate of an epidemic outbreak. 

\par We observe that a defining factor for efficient disease control through voluntary vaccination is the source of information. Higher vaccine coverage is observed for the case when individuals assess their risk of catching infection based on the prevalence in the local social network neighborhood, as opposed to that in the whole population of their social network. The magnitude of global prevalence expressed as the fraction of the entire population size is low in the initial phase of the epidemic, and hence does not appear to pose a severe threat. Consequently, the perception of risk in contracting the disease takes some time to become significant enough to incite vaccine uptake among individuals. However, by the time global prevalence becomes high enough so that the perceived risk of infection outweighs the cost of vaccination, the epidemic will have already affected a large fraction of the population. We find that this delay in the emergence of vaccination behavior can sometimes manifest as a large final size of the epidemic despite high vaccine coverage. On the other hand, the presence of disease in an agent's neighborhood increases the risk of infection even at the early stage of an epidemic, and thus leads to an immediate increase in vaccine uptake. This not only increases the total vaccine coverage but also reduces the burden of disease. An intriguing observation in the case of agents using local information is that the emergence of voluntary vaccination results in a bimodal final epidemic size and vaccine coverage for diseases with $R_0\approx1$. This behavior, observed close to the epidemic threshold, can be attributed to a tight competition between the two possible final outcomes for the state of an initially susceptible individual, namely to get vaccinated or to get infected.

\par Previous game theory based models of vaccination during epidemic outbreaks have considered the effect of strategic decision-making in well-mixed populations where all individuals have the same risk assessment~\cite{Bau2003,Bau2004}. In contrast, our model captures the impact of inhomogeneous risk and benefit perception at the individual level, which gives rise to spatio-temporally diverse games and hence different Nash equilibria across the population. Consequently, the whole population would never converge to a state in which every agent has the same strategy, unless the disease is completely eradicated. This also rules out the possibility that the strategic decision to vaccinate will disappear from the population with time, as suggested by models that utilize imitation game dynamics to describe vaccination behavior~\cite{Fen2011}. These models suggest that the persistence of high vaccine coverage can only be ensured by incentivising vaccine distribution. However, we find that voluntary vaccination can emerge as a response to the potential threat of an epidemic outbreak if each agent utilizes the information available to them and makes a rational decision whether getting vaccinated might be beneficial to her or not. 

\par One of the key assumptions that underpins our approach is that agents are well-informed and make rational decisions based on the information available to them. We have investigated how such rational decision-making affects the final size and vaccine coverage for diseases with different contagiousness (i.e. $R_0$). However, it is possible to augment our model with additional parameters so that it captures the effect of those features of the disease which are qualitatively different from disease contagiousness, such as case fatality ratio. In such a scenario, two diseases with comparable $R_0$, such as Ebola and Influenza, and thus similar transmission rate and vaccination costs, could result in different coverages, based on the subjective perception of how harmful (or severe) a disease is. Another potential extension of our work is to consider the effect of negative sentiments (anti-vaccine sentiments or vaccine scares)~\cite{Lar2011b}, and examine whether the increased chances of catching the infection due to an ongoing epidemic, or witnessing the disease in one's vicinity, can help to close the immunization gap.   

\par We would like to stress that our results are independent of population size and meso-level structural details, such as the existence of modularity, but depend strongly on the degree (average number of contacts a person has) of the network. This could partly be because we are primarily considering the final outcome of the simulated epidemics, such as final epidemic size and total vaccine coverage. Another potential reason is that the strategic decision making in our model depends crucially on the neighborhood which is a micro-level detail of the social network. From a policy-making viewpoint, it is easier to estimate how many social contacts a person has on average rather than meso- and macro-level details, which widens the scope of our model and its results. We also stress on the importance of taking into account the heterogeneity in the disease status of neighbors in a social network for risk assessment when deciding whether to vaccinate. The prevalence aggregated over the whole population may sometimes result in a false perception of risk, especially if the disease is in one's vicinity. The key outcome for public health planning is that accurate and localized reporting of disease outbreak is crucial for changing individuals' risk perception and thereby their attitude towards vaccination, especially during the initial phase of an epidemic.   

\acknowledgments
We would like to thank Kaja Abbas, Samit Bhattacharyya,
Chandrashekar Kuyyamudi, Jose Ruedallano and
Stefan Schuster for helpful discussions.
This work was supported in part by the IMSc Complex Systems Project
($12^{\rm th}$ Plan). The simulations and computations required for
this work were supported by High Performance Computing facility
(Nandadevi) of The Institute of Mathematical Sciences,
which is partially funded by DST.


\end{document}